\documentclass[12pt]{article}
\usepackage{amssymb,amsmath,epsfig}
\allowdisplaybreaks

\begin{document}
\title{\bf Energy Density Inhomogeneities with Polynomial $f(R)$ Cosmology}

\author{M. Sharif \thanks{msharif.math@pu.edu.pk} and Z. Yousaf
\thanks{z.yousaf.math@live.com}\\
Department of Mathematics, University of the Punjab,\\
Quaid-e-Azam Campus, Lahore-54590, Pakistan.}

\date{}

\maketitle
\begin{abstract}
In this paper, we study the effects of polynomial $f(R)$ model on
the stability of homogeneous energy density in self-gravitating
spherical stellar object. For this purpose, we construct couple of
evolution equations which relate the Weyl tensor with matter
parameters. We explore different factors responsible for density
inhomogeneities with non-dissipative dust, isotropic as well as
anisotropic fluids and dissipative dust cloud. We find that shear,
pressure, dissipative parameters and $f(R)$ terms affect the
existence of inhomogeneous energy density.
\end{abstract}
{\bf Keywords:} Dissipative systems; Relativistic systems; Modified
gravity.\\
{\bf PACS:} 04.40.Cv; 04.40.Dg; 04.50.-h.

\section{Introduction}

Recent cosmological evidences predicted by different measurements
indicate transition of our universe from matter dominated epoch to
accelerating expansion state (Perlmutter et al. 1999, Reiss et al.
2007, Komatsu et al. 2011). The accelerating cosmic expansion has
been prompted by an enigmatic ingredient with large negative
pressure, dubbed as dark energy. To explain its nature, different
models like cosmological constant, phantom, quintessence, Chaplygin
gas etc. have been established. The exploration of modified gravity
theories obtained by modifying geometric gravitational part of
Einstein-Hilbert action has received much attention in mathematical
physics. The $f(R)$ gravity (Capozziello 2002, Nojiri and Odintsov
2011) is one of the most viable theories in which Ricci scalar is
replaced by its non-linear generic function. Among important
features of this theory, the likely one is to present a model that
represents early as well as late-time universe expansion in the
absence of dark component. Bamba \textit{et al.} (2012) introduced
unified model for inflation as well as late cosmic expansion model
in this theory. There exist number of $f(R)$ models (Faraoni and
Nadeau 2005, Nojiri and Odintsov 2007, Hu and Sawicki 2007, Bamba et
al. 2012) that correspond to cosmological constraints and pass
experimental test.

Anisotropic pressure in matter configurations results from several
astrophysical factors like pion condensation (Sawyer 1972), various
types of phase transitions (Sokolov 1980), presence of a solid core,
superfluids (Heiselberg and Jensen 2000) as well as strong magnetic
field (Yazadjiev 2012). It is worth stressing that for stable fluid
distribution, anisotropy can endorse outwardly increasing density
within the core of star (Horvat 2011). After the ground work of
Bowers and Liang (1974), there have been number of papers on
pressure anisotropy (Herrera and Santos 1997, Bohmer and Harko 2006,
Herrera et al. 2008, Sharif and Yousaf 2012, Mimoso et al. 2013,
Sharif and Bhatti 2013a, 2013b, 2014b) which assert that anisotropy
may have non-negligible consequences on the structure and properties
of self-gravitating systems. Thirukkanesh and Ragel (2013) presented
spherically symmetric compact star models with anisotropic pressure
which help to understand strange quark stars.

Gravitational collapse is the phenomenon in which massive body falls
inward due to the action of its own gravity that may lead to stars,
star clusters and galaxies from interstellar gas. This occurs due to
extremely inhomogeneous initial state thereby showing the importance
of energy density inhomogeneities in the collapse process. Penrose
(1979) laid much emphasis on the importance of energy density
inhomogeneity in the gravitational time arrow by relating
inhomogeneous density with Weyl tensor. Eardley and Smarr (1979)
asserted that inhomogeneous spherical dust configuration leads to
naked singularities for inhomogeneous collapse. Herrera \textit{et
al.} (1998) discussed the role of density inhomogeneities and local
anisotropy of pressure on the structure and evolution of spherically
symmetric adiabatic self-gravitating objects through the active
gravitational mass. Further, Herrera \textit{et al.} (2004)
investigated density inhomogeneity effects on the evolutionary
phases of dissipative anisotropic spherical systems by evaluating a
link between the Weyl tensor and local anisotropic pressure. Bamba
\textit{et al.} (2011) discussed matter instability and curvature
singularity in the star collapse with $f(R)$ background.

Ziaie \textit{et al.} (2011) studied collapsing mechanism of a star
satisfying barotropic equation of state in $f(R)$ theories and found
finite-time singularities. Borisov \textit{et al.} (2012) analyzed
spherical collapse in metric $f(R)$ gravity with the help of time
evolution numerical simulations. Guo \textit{et al.} (2014) studied
collapse of spherical star in Einstein $f(R)$ frame and concluded
that this may lead to de-Sitter Schwarzschild black hole. We have
investigated impact of late and early time cosmic models on the
collapse of self-gravitating systems with metric as well as Palatini
$f(R)$ theory (Sharif and Yousaf 2013a,b,c,d, 2014a,c).

A great deal of effort has been devoted to study the stability of
stellar systems upon fluctuations. Galli and Koshelev (2011) studied
a class of late-time cosmic evolution models with perturbations
induced by inhomogeneous energy density. Pinheiro and Chan (2011)
examined non-adiabatic anisotropic collapse accompanied by
inhomogeneous density configuration with and without shearing
motion. Sharma and Tikekar (2012) explored shear-free spherical
collapse with dissipation through heat to investigate the
inhomogeneity effects during evolution. Sharif and his collaborators
(Sharif and Yousaf 2012b,c, Sharif and Bhatti 2012a,b, 2014a,c,
Sharif and Tahir 2013) studied spherical, cylindrical and planar
celestial models and analyzed the role of energy density
inhomogeneity in the evolution of fluid parameters that characterize
gravitational collapse.

Herrera \textit{et al.} (2009) explored spherical relativistic fluid
configurations through scalar functions, i.e., $Y_T,~X_T,~Y_{TF}$
and $X_{TF}$ . Herrera \textit{et al.} (2011) extended their results
by invoking cosmological constant to examine the evolution of shear
tensor and expansion scalar. They identified $X_{TF}$ as a factor
describing inhomogeneity in the energy density. Herrera (2011)
discussed stability of inhomogeneous density in anisotropic
spherical fluid configuration with diffusion and free-streaming
approximations. Recently, we have studied dynamics of spherical
matter distribution with structure scalars and $\epsilon R^2$
cosmology (Sharif and Yousaf 2014b).

This paper investigates the role of polynomial $f(R)$ gravity on the
stability of homogeneous energy density with anisotropic and
dissipative spherical matter. The paper is planned as follows. In
section \textbf{2}, we discuss $f(R)$ formulation and relate matter
variables with the Weyl scalar. Section \textbf{3} is devoted to
construct scalar functions with a well-consistent polynomial $f(R)$
model to obtain conservation and Ellis equations. In section
\textbf{4}, we consider various aspects of matter distribution to
analyze density inhomogeneity. In the last section, we conclude our
results.

\section{$f(R)$ Formalism}

The gravitational part of the Einstein-Hilbert action in $f(R)$
gravity is
\begin{equation}\label{1}
S_{f(R)}=\frac{1}{2\kappa}\int d^4x\sqrt{-g}f(R),
\end{equation}
where $\kappa$ and $f(R)$ are coupling constant and a non-linear
generic Ricci scalar function, respectively. The usual GR action can
be retrieved by taking $f(R)=R$. The field equations, in metric
formalism, are calculated by varying Eq.(\ref{1}) with respect to
$g_{\alpha\beta}$ as follows
\begin{equation}\label{2}
R_{\alpha\beta}f_R-\left(\nabla_{\alpha}\nabla_{\beta}-g_{\alpha\beta}{\Box}
\right)f_R-\frac{1}{2}g_{\alpha\beta}f={\kappa}T_{\alpha\beta},
\end{equation}
where $\nabla_\alpha$ and $\Box$ are the covariant derivative and
d'Alembert operator, respectively. Equation (\ref{2}), after some
manipulations, can be expressed as
\begin{equation}\label{3}
G_{\alpha\beta}=\frac{\kappa}{f_R}(\overset{(D)}
{T_{\alpha\beta}}+T_{\alpha\beta}),
\end{equation}
where
\begin{equation*}
\overset{(D)}{T_{\alpha\beta}}=\frac{1}{\kappa}\left\{
\nabla_{\alpha}\nabla_{\beta}f_R+(f-Rf_R)\frac{g_{\alpha\beta}}{2}-\Box
f_Rg_{\alpha\beta}\right\},
\end{equation*}
is the stress energy tensor which indicates gravitational
contribution due to $f(R)$ terms. Under GR limit, i.e.,
$f(R)\rightarrow{R}$, $\overset{(D)}{T_{\alpha\beta}}$ disappears
identically. The system under consideration is modeled as a sphere
with non-static spacetime
\begin{equation}\label{4}
ds^2_-=A^2(t,r)dt^{2}-B^2(t,r)dr^{2}-C^2(t,r)(d\theta^{2}+\sin^2\theta{d\phi^2}),
\end{equation}
consisting of locally anisotropic pressure, dissipating in the
diffusion (heat) and free streaming (null radiation) approximations.
The corresponding stress-energy tensor is
\begin{align}\label{5}
T_{\alpha\beta}=(\mu+P_{\bot})V_\alpha
V_\beta+(P_r-P_\bot)\chi_\alpha\chi_\beta-P_\bot
g_{\alpha\beta}+q_\alpha V_\beta+{\varepsilon}l_\alpha
l_\beta+V_\alpha q_\beta,
\end{align}
where $\mu,~P_{\perp},~P_r,~q_{\beta}$ and $\varepsilon,$ are the
energy density, tangential and radial pressures, heat conducting
vector and radiation density, respectively. Moreover,
$l^\beta,~V^{\beta}$ and $\chi^\beta$ are the null four-vector,
fluid four-velocity and radial unit four-vector, respectively. These
quantities $V^{\beta}=\frac{1}{A}\delta^{\beta}_{0},~
\chi^{\beta}=\frac{1}{B}\delta^{\beta}_{1},~
l^\beta=\frac{1}{A}\delta^{\beta}_{0}+\frac{1}{B}\delta^{\beta}_{1},~
q^\beta=q(t,r)\chi^{\beta}$ in comoving coordinates obey
\begin{eqnarray*}
&&V^{\alpha}V_{\alpha}=-1,\quad\chi^{\alpha}\chi_{\alpha}=1,
\quad\chi^{\alpha}V_{\alpha}=0,\\\nonumber
&&V^\alpha q_\alpha=0, \quad l^\alpha V_\alpha=-1, \quad l^\alpha
l_\alpha=0.
\end{eqnarray*}
The expansion and shear scalars for Eq.(\ref{1}) are given by
\begin{equation}\label{5a}
\Theta{A}=\left(\frac{2\dot{C}}{C}+\frac{\dot{B}}{B}\right), \quad
\sigma{A}=-\left(\frac{\dot{C}}{C} -\frac{\dot{B}}{B}\right),
\end{equation}
where dot stands differentiation with respect to $t$.

The metric $f(R)$ field equations turn out to be
\begin{align}\nonumber
&\frac{\kappa}{f_{R}}\left[A^{2}({\mu}+{\varepsilon})+\frac{A^2}
{\kappa}\left\{\frac{f_{R}'}{B^2}\left(\frac{B'}{B}+\frac{2C'}{C}
\right)-\frac{f_R}{2}\left(R-\frac{f}{f_R}\right)+\frac{f_{R}''}
{B^2}-\frac{\dot{f_{R}}}{A^2}\right.\right.\\\nonumber
&\left.\left.\times\left(\frac{\dot{B}}{B}+\frac{2\dot{C}}{C}
\right)\right\}\right]=\left(\frac{\dot{C}}{C}\right)^2+\frac{
2\dot{C}\dot{B}}{CB}+\left\{\frac{C'}{C}\left(\frac{2B'}{B}-
\frac{C'}{C}\right)+\left(\frac{B}{C}\right)^2\right.\\\label{6}
&\left.-\frac{2C''}{C}\right\}\left(\frac{A}{B}\right)^2,\\\label{7}
&\frac{\kappa}{f_{R}}\left[BA(q+{\varepsilon})-\frac{1}{\kappa}
\left(\dot{f_{R}'}-\frac{\dot{B}f_{R}'}{B}-\frac{A'\dot{f_{R}}}{A}
\right)\right]=2\left(\frac{\dot{C'}}{C}-\frac{A'\dot{C}}{CA}
-\frac{C'\dot{B}}{BC}\right),\\\nonumber
&\frac{\kappa}{f_{R}}\left[B^{2}({P_r}+{\varepsilon})-\frac{B^2}
{\kappa}\left\{\frac{f_{R}'}{B^2}\left(\frac{A'}{A}+\frac{2C'}{C}\right)
-\frac{f_R}{2}\left(R-\frac{f}{f_R}\right)+\left(\frac{\dot{A}}{A}
-\frac{2\dot{C}}{C}\right)\right.\right.\\\label{8}
&\left.\left.\times\frac{\dot{f_{R}}}{A^2}-\frac{\ddot{f_R}}{A^2}
\right\}\right]=\left\{\left(\frac{2\dot{A}}{A}-\frac{\dot{C}}{C}\right)
\frac{\dot{C}}{C}-\frac{2\ddot{C}}{C}\right\}\frac{B^2}{A^2}
-\frac{B^2}{C^2}+\frac{C'}{C}\left(\frac{C'}{C}+\frac{2A'}{A}\right),\\\nonumber
&\frac{\kappa}{f_{R}}\left[{P_{\bot}}C^{2}-\frac{C^2}{\kappa}
\left\{\frac{f_R''}{B^2}-\frac{\ddot{f_{R}}}{A^2}+\left(\frac{
\dot{A}}{A}-\frac{\dot{B}}{B}+\frac{\dot{C}}{C}\right)\frac{
\dot{f_{R}}}{A^2}-\frac{f_R}{2}\left(R-\frac{f}{f_R}\right)\right.\right.\\\nonumber
&\left.+\left(\frac{C'}{C}-\frac{B'}{B}\left.+\frac{A'}{A}\right)
\frac{f_{R}'}{B^2}\right\}\right]=\left\{\frac{\dot{C}}{C}
\left(\frac{\dot{A}}{A}-\frac{\dot{B}}{B}\right)-\frac{\ddot{C}}{C}
+\frac{\dot{B}\dot{A}}{BA}-\frac{\ddot{B}}{B}\right\}\frac{C^2}{A^2}
\\\label{9}
&+\left\{\frac{A'}{A}\left(\frac{C'}{C}-\frac{B'}{B}\right)
\frac{C''}{C}-\frac{B'C'}{BC}+\frac{A''}{A}\right\}\frac{C^2}{B^2},
\end{align}
where prime represents differentiation with respect to $r$ . The
Misner-Sharp mass function is given by (Misner and Sharp 1964)
\begin{equation}\label{10}
m(t,r)=\frac{C}{2}(1-g^{\alpha\beta}C_{,\alpha}C_{,\beta})
=\left\{1+\left(\frac{\dot{C}}{A}\right)^2
-\left(\frac{C'}{B}\right)^2\right\}\frac{C}{2}.
\end{equation}
The radial and proper derivative operators are defined respectively
as follows
\begin{eqnarray}\label{11}
D_{C}=\frac{1}{C'}\frac{\partial}{\partial r},\quad
D_{T}=\frac{1}{A} \frac{\partial}{\partial t}.
\end{eqnarray}
The proper time rate of change of areal radius of the spherical
system is
\begin{eqnarray}\label{12}
U=D_{T}C=\frac{\dot{C}}{A}<1~\textmd{(for collapsing bodies)}.
\end{eqnarray}
In terms of collapsing fluid velocity, Eq.(\ref{10}) can be written
as
\begin{eqnarray}\label{13}
E\equiv\frac{C'}{B}=\left[1+U^{2}-\frac{2m(t,r)}{C}\right]^{1/2}.
\end{eqnarray}

The time and radial mass variations can be followed from
Eqs.(\ref{6})-(\ref{8}), (\ref{10}) and (\ref{11}) as
\begin{align}\label{14}
&D_{T}m=-\frac{\kappa}{2f_{R}}\left\{U\left(\hat{P_r}
+\frac{\overset{~~~(D)}{T_{11}}}{B^2}\right)+E\left(\hat{q}
-\frac{\overset{~~~(D)}{T_{01}}}{AB}\right)\right\}C^2,\\\label{15}
&D_{C}m=\frac{\kappa}{2f_{R}}\left\{\hat{\mu}+\frac{
\overset{~~~(D)}{T_{00}}}{A^2}+\frac{U}{E}\left(\hat{q}
-\frac{\overset{~~~(D)}{T_{01}}}{AB}\right)\right\}C^2,
\end{align}
where $\hat{P_r}=P_r+{\varepsilon},~\hat{q}=q+{\varepsilon}$ and
$\hat{\mu}=\mu+{\varepsilon}$. Integration of Eq.(\ref{15}) provides
\begin{equation}\label{16}
\frac{3m}{C^3}=\frac{3\kappa}{2C^3}\int^r_{0}\left[\frac{1}{f_{R}}
\left\{\hat{\mu}+\frac{\overset{~~~(D)}{T_{00}}}{A^2}
+\left(\hat{q}-\frac{\overset{~~~(D)}{T_{01}}}{AB}\right)
\frac{U}{E}\right\}C^2C'\right]dr,
\end{equation}
thereby relating mass function and other fluid variables with $f(R)$
terms. The electric component of the Weyl tensor in terms of
$\chi_\alpha$ and unit four velocity is given by
\begin{equation*}\nonumber
E_{\alpha\beta}=\mathcal{E}\left[\chi_{\alpha}\chi_{\beta}-\frac{1}{3}
(g_{\alpha\beta}+V_\alpha V_\beta)\right],
\end{equation*}
where
\begin{align}\nonumber
\mathcal{E}&=\left[\frac{\ddot{C}}{C}+\left(\frac{\dot{B}}{B}
-\frac{\dot{C}}{C}\right)\left(\frac{\dot{C}}{C}+\frac{
\dot{A}}{A}\right)-\frac{\ddot{B}}{B}\right]\frac{1}{2A^{2}}
-\frac{1}{2C^{2}}\\\label{17}
&-\left[\frac{C''}{C}-\left(\frac{C'}{C}+\frac{B'}{B}\right)
\left(\frac{A'}{A}-\frac{C'}{C}\right)-\frac{A''}{A}\right]\frac{1}{2B^{2}},
\end{align}
which after using Eqs.(\ref{6}) and (\ref{8})-(\ref{10}) can be
expressed as
\begin{align}\label{18}
\frac{3m}{C^3}&=\frac{\kappa}{2f_R}\left(\hat{\mu}-\hat{\Pi}
+\frac{\overset{~~~(D)}{T_{00}}}{A^2}-\frac{\overset{~~~(D)}{T_{11}}}{B^2}
+\frac{\overset{~~~(D)}{T_{22}}}{C^2}\right)-\mathcal{E},
\end{align}
where $\hat{\Pi}=\hat{P}_r-P_\bot$. This equation peculiarly relates
mass function with Weyl scalar and all the fluid variables in $f(R)$
gravity.

\section{Structure Scalars and Ellis Equations}

In this section, we first discuss a viable $f(R)$ model and then
construct structure scalars. We also write down conservation laws
from the usual as well as effective stress energy tensors and
develop the so called Ellis equations. We take a polynomial
inflationary model given as follows (Huang 2014)
\begin{equation}\label{19}
f(R)=R+{\epsilon}R^{2}+\frac{\lambda_n(2{\epsilon}R)^n}{4n\epsilon},
\end{equation}
where $\epsilon=\frac{1}{6M^2}$ and $\lambda_n$ is a dimension-free
coupling parameter with $n>2$. Here energy scale $M$ is refined in
order to make unit normalization to the higher coefficient of $R^2$
term. This model corresponds to the model with
$f(R)=R+R^n/(3M^2)^n-1$ under $\lambda_n\gg1$ while
$\lambda_n\rightarrow0$ provides Starobinsky model (1980). In the
limit $\lambda_n\ll1,~R^n$ terms serve as a small correction to the
inflationary $R+\epsilon R^2$ model which of course makes the model
expansion around the Starobinsky model. It is interesting to mention
here that the inflation induced by $R+R^4$ gravity provides much
different platform than that of $R+R^2$ gravity and is close to
topological inflation (Saidov 2010). All GR solutions can be found
by taking limit $f(R)\rightarrow R$.

To formulate $f(R)$ structure scalars, we orthogonally split the
Riemann tensor and propose tensors $X_{\alpha\beta}$ and
$Y_{\alpha\beta}$ as (Herrera et al. 2011)
\begin{equation*}
X_{\alpha\beta}=^{*}R^{*}_{\alpha\mu\beta\nu}V^{\mu}V^{\nu}=
\frac{1}{2}\eta^{\epsilon\rho}_{~~\alpha\mu}R^{*}_{\epsilon
\rho\beta\nu}V^{\mu}V^{\nu},\quad
Y_{\alpha\beta}=R_{\alpha\mu\beta\nu}V^{\mu}V^{\nu},
\end{equation*}
where $R^{*}_{\alpha\beta\mu\nu}=
\frac{1}{2}\eta_{\epsilon\rho\mu\nu}R^{\epsilon\rho}_{~~\alpha
\beta}$. These equations in terms of trace and trace-less components
are given by
\begin{align}\label{20}
X_{\alpha\beta}&=\frac{1}{3}X_{T}h_{\alpha\beta}
+X_{TF}\left(\chi_{\alpha}\chi_{\beta}-\frac{1}{3}h_{\alpha\beta}
\right),\\\label{21}
Y_{\alpha\beta}&=\frac{1}{3}Y_{T}h_{\alpha\beta}+Y_{TF}\left(
\chi_{\alpha}\chi_{\beta} -\frac{1}{3}h_{\alpha\beta}\right).
\end{align}
We use Eqs.(\ref{6}), (\ref{8}), (\ref{9}) and (\ref{19})-(\ref{21})
with some manipulations to obtain the following scalar structures
\begin{align}\label{22}
&X_{T}=\frac{4\kappa{\epsilon}R}{4{\epsilon}R(1+2{\epsilon}R)
+\lambda_n(2{\epsilon}R)^n}\left(\hat{\mu}+\frac{\varphi_\mu}
{A^2}\right),\\\label{23}
&X_{TF}=-\mathcal{E}-\frac{2\kappa{\epsilon}R}{4{\epsilon}
R(1+2{\epsilon}R)+\lambda_n(2{\epsilon}R)^n}
\left(\hat{\Pi}-2{\sigma}{\eta}+\frac{\varphi_{P_r}}{B^2}
-\frac{\varphi_{P_\bot}}{C^2}\right),\\\label{24}
&Y_{T}=\frac{2\kappa{\epsilon}R}{4{\epsilon}R(1+2{\epsilon}R)
+\lambda_n(2{\epsilon}R)^n}\left(\hat{\mu}+\frac{\varphi_\mu}
{A^2}+\frac{\varphi_{P_r}}{B^2}+\frac{2\varphi_{P_\bot}}
{C^2}+3\hat{P_{r}}-2\hat{\Pi}\right),\\\label{25}
&Y_{TF}=\mathcal{E}-\frac{2\kappa{\epsilon}R}{4{\epsilon}R(1
+2{\epsilon}R)+\lambda_n(2{\epsilon}R)^n}\left(\hat{\Pi}-2{\eta}
{\sigma}+\frac{\varphi_{P_r}}{B^2}-\frac{\varphi_{P_\bot}}{C^2}\right),
\end{align}
where $\varphi_{\mu},~\varphi_{P_r}$ and $\varphi_{P_\bot}$ are
given in Appendix \textbf{A}. It is well established in GR as well
as in $f(R)$ gravity that, one of the structure scalars $X_T$
describes matter energy density while its inhomogeneity is discussed
with the help of $X_{TF}$ only if the system evolves adiabatically
alongwith $\epsilon=0$. The scalar functions $Y_{TF}$ and $Y_{T}$
incorporating $\epsilon$ terms control the evolutionary mechanisms
of shearing and expansion rates of the system.

The two independent components of the contracted Bianchi identities
are
\begin{align}\label{26}
\left(\overset{(D)}
{T^{\alpha\beta}+T^{\alpha\beta}}\right)_{;\beta}=0,\quad
\left(\overset{(D)}
{T^{\alpha\beta}+T^{\alpha\beta}}\right)_{;\beta}=0,
\end{align}
which yield
\begin{align}\label{27}
&\hat{\dot{\mu}}+\frac{A\hat{q}'}{B}+(\hat{P}_r+\hat{\mu})
\frac{\dot{B}}{B}+\frac{2A\hat{q}C'}{BC}+2(P_\bot+\hat{\mu})
\frac{\dot{C}}{C}+D_0(t,r)=0,\\\label{28}
&\frac{A\hat{{P'_r}}}{B}+\dot{\hat{q}}+(\hat{P}_r+\hat{\mu})
\frac{A'}{B}+2\left(\frac{\dot{C}}{C}+\frac{\dot{B}}{B}\right)
\hat{q}+2\hat{\Pi}\frac{(AC)'}{BC}+D_1(t,r)=0,
\end{align}
where $D_0$ and $D_1$ are $f(R)$ dark source terms given in Appendix
\textbf{A}. Now we find two very important differential equations
which play a pivotal in the stability analysis of inhomogeneous
energy density. These two equations were firstly calculated by Ellis
(2009) and then by Herrera \textit{et al.} (2004) in GR. These
equations are obtained by using Eqs.(\ref{6})-(\ref{9}), (\ref{14}),
(\ref{15}) and (\ref{19}) as
\begin{align}\nonumber
&\left[\mathcal{E}-\frac{2\kappa{\epsilon}R}{4{\epsilon}R(1+2{\epsilon}R)
+\lambda_n(2{\epsilon}R)^n}\left(\hat{\mu}-\hat{\Pi}+\frac{\varphi_\mu}
{A^2}-\frac{\varphi_{P_r}}{B^2}+\frac{\varphi_{P_\bot}}{C^2}\right)
\right]_{,0}\\\nonumber
&=\frac{3\dot{C}}{C}\left[\frac{2\kappa{\epsilon}R}{4{\epsilon}R
(1+2{\epsilon}R)+\lambda_n(2{\epsilon}R)^n}\left(\hat{\mu}+\hat
{P_\bot}+\frac{\varphi_\mu}{A^2}+\frac{\varphi_{P_\bot}}{C^2}
\right)-\mathcal{E}\right]\\\label{29}
&+\frac{6\kappa{\epsilon}R}{4{\epsilon}R(1+2{\epsilon}R)
+\lambda_n(2{\epsilon}R)^n}\left(\frac{AC'}{BC}\right)
\left(\hat{q}-\frac{\varphi_q}{AB}\right),\\\nonumber
&\left[\mathcal{E}-\frac{2\kappa{\epsilon}R}{4{\epsilon}
R(1+2{\epsilon}R)+\lambda_n(2{\epsilon}R)^n}\left(\hat{\mu}-\hat
{\Pi}+\frac{\varphi_\mu}{A^2}-\frac{\varphi_{P_r}}{B^2}
+\frac{\varphi_{P_\bot}}{C^2}\right)\right]_{,1}\\\nonumber
&=-\frac{3{C'}}{C}\left[\frac{2\kappa{\epsilon}R}
{4{\epsilon}R(1+2{\epsilon}R)+\lambda_n(2{\epsilon}R)^n}
\left(\hat{\mu}+\frac{\varphi_\mu}{A^2}\right)-\frac{3m}
{C^3}\right]\\\label{30}
&-\frac{6\kappa{\epsilon}R}{4{\epsilon}R(1+2{\epsilon}R)
+\lambda_n(2{\epsilon}R)^n}\left(\frac{B\dot{C}}{AC}\right)
\left(\hat{q}-\frac{\varphi_q}{AB}\right),
\end{align}
where $\varphi_q$ is mentioned in Appendix \textbf{A}. Both of the
above equations reduce to GR (Herrera 2011) under
$\epsilon\rightarrow0$.

\section{Stability of Homogeneous Energy Density}

In this section, we discuss different factors affecting energy
density homogeneity in matter distribution with $f(R)$ framework for
different cases. We confine ourselves with present valued
cosmological Ricci scalar, i.e., $R=\tilde{R}$.

\subsection{Non-dissipative Fluids}

In this subsection, we perform our analysis with non-dissipative
matter distribution with polynomial $f(R)$ gravity model for dust,
isotropic and anisotropic fluid configurations.

\subsubsection{Dust Cloud}

Here we take non-dissipative dust fluid with its geodesic motion
which gives $\hat{q}=P_\bot=\hat{P}_r=0$ and $A=1$. In this context,
Eqs.(\ref{29}) and (\ref{30}) give
\begin{align}\nonumber
&\left[\mathcal{E}-\frac{2\kappa{\epsilon}\tilde{R}}{4{\epsilon}
\tilde{R}(1+2{\epsilon}\tilde{R})+\lambda_n(2{\epsilon}\tilde{R})^n}
\left({\mu}+\frac{\lambda_n(1-n)(2{\epsilon}\tilde{R})^n}{8\kappa{\epsilon}n}
-\frac{{\epsilon}\tilde{R}^2}{2\kappa}\right)\right]_{,0}\\\label{31}
&=\frac{3\dot{C}}{C}\left[\frac{2\kappa{\epsilon}\tilde{R}{\mu}}
{4{\epsilon}\tilde{R}(1+2{\epsilon}\tilde{R})+\lambda_n(2{\epsilon}
\tilde{R})^n}-\mathcal{E}\right],\\\nonumber
&\left[\mathcal{E}-\frac{2\kappa{\epsilon}\tilde{R}}{4{\epsilon}
\tilde{R}(1+2{\epsilon}\tilde{R})+\lambda_n(2{\epsilon}\tilde{R})^n}
\left({\mu}+\frac{\lambda_n(1-n)(2{\epsilon}\tilde{R})^n}{8\kappa{\epsilon}n}
-\frac{{\epsilon}\tilde{R}^2}{2\kappa}\right)\right]'\\\label{32}
&=-\frac{3{C'}}{C}\mathcal{E}.
\end{align}
By making use of Eqs.(\ref{5a}), (\ref{27}) and
(\ref{B3})-(\ref{B6}) in Eqs.(\ref{31}) and (\ref{32}), we obtain
\begin{align}\label{33a}
&\dot{\mathcal{E}}+\frac{3\dot{C}}{C}\mathcal{E}=\frac{-2\kappa
{\epsilon}A\sigma\mu\tilde{R}}{4{\epsilon}\tilde{R}(1+2{\epsilon}\tilde{R})
+\lambda_n(2{\epsilon}\tilde{R})^n},\\\label{33b}
&{\mathcal{E}}'+\frac{3{C'}}{C}\mathcal{E}=\frac{2\kappa{\epsilon}
\tilde{R}\mu'}{4{\epsilon}\tilde{R}(1+2{\epsilon}\tilde{R})
+\lambda_n(2{\epsilon}\tilde{R})^n}.
\end{align}
Equation (\ref{33a}) yields
\begin{align}\label{33}
&{\mathcal{E}}=\frac{-2\kappa{\epsilon}
\tilde{R}\int_0^tA\sigma{\mu}C^3dt}{[4{\epsilon}\tilde{R}(1+2{\epsilon}\tilde{R})
+\lambda_n(2{\epsilon}\tilde{R})^n]C^3},
\end{align}
which provides condition for the existence of homogeneity in the
dust fluid. This states that non-dissipative homogeneous spherical
matter configuration exists only if the system is conformally flat.
Similarly, we can identify the Weyl scalar as an inhomogeneity
factor from Eq.(\ref{33b}). Equation (\ref{33}) also asserts that
conformal flatness exists if the system evolves with vanishing shear
scalar.

\subsubsection{Isotropic Fluid}

Now, we consider adiabatic spherical system with locally isotropic
pressure. Under this scenario, Eqs.(\ref{29}) and (\ref{30}) become
\begin{align}\nonumber
&\left[\mathcal{E}-\frac{2\kappa{\epsilon}\tilde{R}}{4{\epsilon}
\tilde{R}(1+2{\epsilon}\tilde{R})+\lambda_n(2{\epsilon}\tilde{R})^n}
\left({\mu}+\frac{\lambda_n(1-n)(2{\epsilon}\tilde{R})^n}{8\kappa{\epsilon}n}
-\frac{{\epsilon}\tilde{R}^2}{2\kappa}\right)\right]_{,0}\\\label{34}
&+\frac{3\dot{C}}{C}\left[\frac{-2\kappa{\epsilon}\tilde{R}({\mu}+P)}
{4{\epsilon}\tilde{R}(1+2{\epsilon}\tilde{R})+\lambda_n(2{\epsilon}
\tilde{R})^n}+\mathcal{E}\right]=0,\\\nonumber
&\left[\mathcal{E}-\frac{2\kappa{\epsilon}\tilde{R}}{4{\epsilon}\tilde{R}
(1+2{\epsilon}\tilde{R})+\lambda_n(2{\epsilon}\tilde{R})^n}\left({\mu}
+\frac{\lambda_n(1-n)(2{\epsilon}\tilde{R})^n}{8\kappa{\epsilon}n}
-\frac{{\epsilon}\tilde{R}^2}{2\kappa}\right)\right]'\\\label{35}
&+\frac{3{C'}}{C}\mathcal{E}=0.
\end{align}
It is seen that Eq.(\ref{35}) turns out to be same as Eq.(\ref{32}),
thus showing that $\mathcal{E}=0$ if and only if $\mu'=0$. Equation
(\ref{34}) after using Eqs.(\ref{5a}) and (\ref{27}) provides the
following differential equation
\begin{align}\label{36a}
&\dot{\mathcal{E}}+\frac{3\dot{C}}{C}\mathcal{E}=\frac{-2\kappa{
\epsilon}A\sigma(\mu+P)\tilde{R}}{4{\epsilon}\tilde{R}(1+2{\epsilon}
\tilde{R})+\lambda_n(2{\epsilon}\tilde{R})^n},
\end{align}
whose solution is
\begin{align}\label{36}
&{\mathcal{E}}=\frac{-2\kappa{\epsilon}\tilde{R}\int_0^tA\sigma({\mu}+P)
C^3dt}{[4{\epsilon}\tilde{R}(1+2{\epsilon}\tilde{R})
+\lambda_n(2{\epsilon}\tilde{R})^n]C^3}.
\end{align}
This argues that energy density of the system will be homogeneous as
long as the system embodies shear-free motion. Thus the condition of
locally isotropic pressure in the matter configuration with constant
Ricci scalar $f(R)$ model does not affect the stability of
homogeneous energy density found in the above case. Let us assume
shear-free fluid so that Eq.(\ref{36a}) provides
\begin{align*}\nonumber
&\mathcal{E}=\frac{\omega(r)}{C^3},
\end{align*}
where $\omega(r)$ is an integration function. If the system is
homogeneous initially, i.e., $\mathcal{E}(0,r)=0$, then $\omega=0$
yields $\mathcal{E}(t,r)=0$. Thus the above condition for
homogeneous energy density will be valid from $t=0$ to onward.
However, if the fluid is expanding such that $\mathcal{E}$ has very
small (non-zero) value at the initial stage, then it will remain as
it is in all the evolutionary phases. If instead the system is
contracting, then the Weyl scalar does not vanish for all time.

\subsubsection{Anisotropic Fluid}

This case corresponds to anisotropic but non-dissipating matter
distribution, i.e. $\Pi\neq0$ and $\hat{q}=0$. In this framework,
Eqs.(\ref{29}) and (\ref{30}) provide
\begin{align}\nonumber
&\left[\mathcal{E}-\frac{2\kappa{\epsilon}\tilde{R}}{4{\epsilon}
\tilde{R}(1+2{\epsilon}\tilde{R})+\lambda_n(2{\epsilon}\tilde{R})^n}
\left({\mu}-\Pi+\frac{\lambda_n(1-n)(2{\epsilon}\tilde{R})^n}{8\kappa{\epsilon}n}
-\frac{{\epsilon}\tilde{R}^2}{2\kappa}\right)\right]_{,0}\\\label{37}
&=\frac{3\dot{C}}{C}\left[\frac{2\kappa{\epsilon}\tilde{R}({\mu}+P_\bot)}
{4{\epsilon}\tilde{R}(1+2{\epsilon}\tilde{R})+\lambda_n(2{\epsilon}
\tilde{R})^n}-\mathcal{E}\right],\\\nonumber
&\left[\mathcal{E}-\frac{2\kappa{\epsilon}\tilde{R}}{4{\epsilon}
\tilde{R}(1+2{\epsilon}\tilde{R})+\lambda_n(2{\epsilon}\tilde{R})^n}
\left({\mu}-\Pi+\frac{\lambda_n(1-n)(2{\epsilon}\tilde{R})^n}{8\kappa{\epsilon}n}
-\frac{{\epsilon}\tilde{R}^2}{2\kappa}\right)\right]'\\\label{38}
&=-\frac{3{C'}}{C}\left[\mathcal{E}+\frac{2\kappa{\epsilon}\tilde{R}
\Pi}{4{\epsilon}\tilde{R}(1+2{\epsilon}\tilde{R})
+\lambda_n(2{\epsilon}\tilde{R})^n}\right].
\end{align}
We use Eqs.(\ref{5a}) and (\ref{27}) in Eqs.(\ref{37}) and
(\ref{38}) to obtain the following set of evolutionary equations
\begin{align}\nonumber
&\left[\mathcal{E}+\frac{2\kappa{\epsilon}\tilde{R}\Pi}{4{\epsilon}
\tilde{R}(1+2{\epsilon}\tilde{R})+\lambda_n(2{\epsilon}\tilde{R})^n}
\right]_{,0}+\frac{3\dot{C}}{C}\left[\mathcal{E}+\frac{2\kappa{\epsilon}
\tilde{R}\Pi}{4{\epsilon}\tilde{R}(1+2{\epsilon}\tilde{R})+\lambda_n
(2{\epsilon}\tilde{R})^n}\right]\\\nonumber
&=\frac{-2\kappa{\epsilon}\tilde{R}}{4{\epsilon}\tilde{R}(1+2{\epsilon}
\tilde{R})+\lambda_n(2{\epsilon}\tilde{R})^n}\left\{A\sigma(\mu+P_r)
-2\Pi\frac{\dot{C}}{C}\right\},\\\nonumber
&\left[\mathcal{E}+\frac{2\kappa{\epsilon}\tilde{R}\Pi}{4{\epsilon}
\tilde{R}(1+2{\epsilon}R)+\lambda_n(2{\epsilon}\tilde{R})^n}\right]'
-\frac{2\kappa{\epsilon}\tilde{R}\mu'}{4{\epsilon}\tilde{R}(1+2{\epsilon}
\tilde{R})+\lambda_n(2{\epsilon}\tilde{R})^n}\\\nonumber
&=-\frac{3{C'}}{C}\left[\mathcal{E}+\frac{2\kappa{\epsilon}\tilde{R}
\Pi}{4{\epsilon}\tilde{R}(1+2{\epsilon}\tilde{R})+\lambda_n(2{\epsilon}
\tilde{R})^n}\right].
\end{align}

These equations in terms of structure scalar (\ref{23}) can be
written as
\begin{align}\nonumber
&\dot{X}_{TF}+\frac{3X_{TF}\dot{C}}{C}=\frac{2\kappa{\epsilon}
\tilde{R}}{4{\epsilon}\tilde{R}(1+2{\epsilon}\tilde{R})+\lambda_n
(2{\epsilon}\tilde{R})^n}\left\{A\sigma(\mu+P_r)-2\Pi\frac{\dot{C}}
{C}\right\},\\\nonumber
&X'_{TF}+\frac{3X_{TF}{C'}}{C}=-\frac{2\kappa{\epsilon}\tilde{R}
\mu'}{4{\epsilon}\tilde{R}(1+2{\epsilon}\tilde{R})+\lambda_n(2{\epsilon}
\tilde{R})^n},
\end{align}
whose general solutions can be found respectively as
\begin{align}\label{39}
&X_{TF}=-\frac{2\kappa{\epsilon}\tilde{R}\int_0^t[2\Pi\dot{C}
-AC\sigma({\mu}+P_r)]C^2dt}{C^3[4{\epsilon}\tilde{R}(1+2{\epsilon}
\tilde{R})+\lambda_n(2{\epsilon}\tilde{R})^n]},\\\label{40}
&X_{TF}=-\frac{2\kappa{\epsilon}\tilde{R}\int_0^rC^3\mu'dr}
{C^3[4{\epsilon}\tilde{R}(1+2{\epsilon}\tilde{R})+\lambda_n
(2{\epsilon}\tilde{R})^n]}.
\end{align}
These equations indicate that quantity incorporating stability of
inhomogeneous energy density is one of the $f(R)$ structure scalars,
i.e., $X_{TF}$. Equation (\ref{40}) shows that  $\mu'=0$ if and only
if $X_{TF}$ vanishes, thereby showing $X_{TF}$ as a factor of
controlling inhomogeneity in anisotropic spherical system which is
well-consistent with (Herrera et al. 2009, Herrera et al. 2011,
Herrera 2011). Thus the inclusion of dark matter/energy effects in
the evolving system do not disrupt the importance of $X_{TF}$. Also,
the above expressions reduce to GR under the limit
$\epsilon\rightarrow0$. However, Eq.(\ref{39}) asserts that
anisotropic pressure, $f(R)$ model and shear scalar are responsible
for the emergence of inhomogeneous energy density in the matter
distribution.
\subsection{Dissipative Dust Cloud}

To see the effects of radiation density and heat conducting vector
in the inhomogeneous energy density, we assume geodesic case, i.e.,
$P_r=P_\bot=0$ with $A=1$. Many authors (Herrera et al. 2004,
Herrera 2011, Kolassis et al. 1988, Govender et al. 1998,
Thirukkanesh and Maharaj 2009, Naidu et al. 2006) discussed
spherical dissipative collapsing dust models with geodesics in order
to explore dissipation effects through the system. In this case,
Eqs.(\ref{29}) and (\ref{30}) yield
\begin{align}\nonumber
&\left[\mathcal{E}-\frac{2\kappa{\epsilon}\tilde{R}}
{4{\epsilon}\tilde{R}(1+2{\epsilon}\tilde{R})+\lambda_n
(2{\epsilon}\tilde{R})^n}\left({\mu}+\frac{\lambda_n
(1-n)(2{\epsilon}\tilde{R})^n}{8\kappa{\epsilon}n}
-\frac{{\epsilon}\tilde{R}^2}{2\kappa}\right)\right]_{,0}
=\frac{3\dot{C}}{C}\\\label{41}
&\times\left[\frac{2\kappa{\epsilon}\tilde{R}{\mu}}
{4{\epsilon}\tilde{R}(1+2{\epsilon}\tilde{R})+\lambda_n
(2{\epsilon}\tilde{R})^n}-\mathcal{E}\right]+\frac{2
\kappa{\epsilon}\tilde{R}}{4{\epsilon}\tilde{R}(1+2
{\epsilon}\tilde{R})+\lambda_n(2{\epsilon}\tilde{R})^n}
\left(\frac{AC'\hat{q}}{BC}\right),\\\nonumber
&\left[\mathcal{E}-\frac{2\kappa{\epsilon}\tilde{R}}
{4{\epsilon}\tilde{R}(1+2{\epsilon}\tilde{R})+\lambda_n
(2{\epsilon}\tilde{R})^n}\left({\mu}+\frac{\lambda_n
(1-n)(2{\epsilon}\tilde{R})^n}{8\kappa{\epsilon}n}
-\frac{{\epsilon}\tilde{R}^2}{2\kappa}\right)\right]'\\\label{42}
&=-\frac{3{C'}}{C}\mathcal{E}-\frac{6\kappa{\epsilon}
\tilde{R}}{4{\epsilon}\tilde{R}(1+2{\epsilon}\tilde{R})
+\lambda_n(2{\epsilon}\tilde{R})^n}\left(\frac{B\hat{q}
\dot{C}}{AC}\right).
\end{align}
Equation (\ref{42}) gives
\begin{align}\label{43}
&\Phi\equiv\mathcal{E}-\frac{6\kappa{\epsilon}\tilde{R}\int^r_0BC^2
\tilde{q}\dot{C}dr}{4{\epsilon}\tilde{R}(1+2{\epsilon}\tilde{R})
+\lambda_n(2{\epsilon}\tilde{R})^n}.
\end{align}
It is found that $\mu'=0$ if and only if $\Phi=0$, indicating that
$\Phi$ is responsible for fluid density inhomogeneity in the dust
spherical system with free streaming and diffusion approximations.
We use Eqs.(\ref{5a}) and (\ref{27}) in Eq.(\ref{41}) to obtain
$\Phi$ evolution equation as follows
\begin{align}\label{44}
&\dot{\Phi}-\frac{\dot{\Psi}}{C^3}=\frac{2\kappa{\epsilon}
\tilde{R}}{4{\epsilon}\tilde{R}(1+2{\epsilon}\tilde{R})
+\lambda_n(2{\epsilon}\tilde{R})^n}\left(\frac{\tilde{q}C'}{BC}
-\tilde{\mu}\sigma
-\frac{\tilde{q}'}{B}\right)-\frac{3\dot{C}}{C}\Phi,
\end{align}
with $\Psi=\frac{6\kappa{\epsilon}
\tilde{R}}{4{\epsilon}\tilde{R}(1+2{\epsilon}\tilde{R})
+\lambda_n(2{\epsilon}\tilde{R})^n}\int^r_0BC^2\tilde{q}\dot{C}dr$,
which yields $\Phi$ as follows
\begin{align}\label{45}
&\Phi=\frac{\int^t_0\left[\dot{\Psi}+\frac{2\kappa{\epsilon}C^2
\tilde{R}}{4{\epsilon}\tilde{R}(1+2{\epsilon}\tilde{R})
+\lambda_n(2{\epsilon}\tilde{R})^n}\left(\frac{\tilde{q}C'}{B}
-\tilde{\mu}C\sigma -\frac{\tilde{q}'C}{B}\right)\right]dt}{C^3}.
\end{align}
This indicates that various fluid parameters affect the evolution of
$\Phi$ in the self-gravitating system. We also see from the above
relation that existence of inhomogeneous density depends upon two
factors, i.e., dissipative parameters and shear scalar. This
describes that shearing scalar, radiation density and heat
dissipation hold fundamental importance in the the study of
inhomogeneous matter distribution leading to gravitational collapse.

\section{Summary and Discussion}

This work analyzes various factors producing inhomogeneity in the
energy density of the spherical self-gravitating celestial body in
$f(R)$ gravity. We have constructed structure scalars by
orthogonally splitting the Riemann curvature tensor to obtain
evolution equations using a viable inflationary $f(R)$ model. We
have discussed our analysis for non-dissipative dust, isotropic as
well as anisotropic fluid configurations and dust cloud dissipating
fluid. The results are concluded as follows.
\begin{itemize}
\item For non-dissipative dust and locally isotropic ideal matter cloud,
it is seen from Eqs.(\ref{33b}) and (\ref{35}) that the system will
encapsulate homogeneous energy density if and only if the system is
conformally flat. The extra $f(\tilde{R})$ degrees of freedom terms
turn down contribution of $\mathcal{E}$, thus relaxing conformal
flatness condition.
\item In an adiabatic anisotropic spherical system, the density inhomogeneity
is described in terms of pressure anisotropy which in turn
controlled by one of the structure scalars, $X_{TF}$ as mentioned in
Eq.(\ref{23}). Equation (\ref{40}) also establishes $X_{TF}$ as an
element of governing inhomogeneity in the system. This result is
well-consistent with (Sharif and Yousaf 2014b) under
$\lambda_n\rightarrow0$ and (Herrera 2011) under
$\epsilon\rightarrow0$ which correspond to solutions in $R+\epsilon
R^2$ gravity and GR, respectively.
\item The quantity $\Phi$ is explored and identified to be responsible
for the emergence of inhomogeneity in energy density for geodesic
radiating dust fluid. Extra curvature $f(R)$ terms, dissipation
parameters and shear scalar affect evolution of $\Phi$ as described
by Eq.(\ref{45}).
\item All these results correspond to GR under $\epsilon\rightarrow0$ (Herrera 2011).
It is worth stressing that structure scalars obtained in
Eqs.(\ref{22})-(\ref{25}) hold fundamental importance in the study
of self-gravitating system. For $\lambda_n\rightarrow0$, scalar
functions reduces for $f(R)=R+\epsilon R^2$ cosmology (Sharif and
Yousaf 2014b) while $\epsilon\rightarrow0$ provides GR results
(Herrera et al. 2011).
\end{itemize}

\vspace{0.3cm}

\renewcommand{\theequation}{B\arabic{equation}}
\setcounter{equation}{0}
\section*{Appendix A}

The higher curvature terms $D_0$ and $D_1$ of Eqs.(\ref{27}) and
(\ref{28}) are given as
\begin{align}\nonumber
D_0&=\frac{1}{\kappa}\left[\left\{\left(\frac{A'}{A}\dot{f_R}
+\frac{\dot{B}}{B}f'_R-\dot{f'_R}\right)\frac{1}{A^2B^2}\right\}_{,1}
+\frac{1}{A^2}\left\{\frac{f_R}{2}\left(\frac{f}{f_R}-R\right)\right.\right.\\\nonumber
&+\left.\left.\left(\frac{2C'}{C}+\frac{B'}{B}\right)\frac{f'_R}{B^2}
+\frac{f''_R}{B^2}-\frac{\dot{f_R}}{A^2}\left(\frac{2\dot{C}}{C}
+\frac{\dot{B}}{B}\right)\right\}_{,0}+\frac{1}{A^2}\left\{\frac{
\ddot{f_R}}{A^2}\right.\right.\\\nonumber
&\left.\left.-\left(\frac{B'}{B}+\frac{A'}{A}\right)\frac{f'_R}{B^2}
+\frac{f''_R}{B^2}-\left(\frac{\dot{B}}{B}+\frac{\dot{A}}{A}\right)
\frac{\dot{f_R}}{A^2}\right\}\frac{\dot{B}}{B}+\frac{2}{A^2}\left\{
-\left(\frac{A'}{A}\right.\right.\right.\\\nonumber
&\left.\left.\left.-\frac{C'}{C}\right)\frac{f'_R}{B^2}+\frac{\ddot{f_R}}
{A^2}-\frac{\dot{f_R}}{A^2}\left(\frac{3\dot{C}}{C}+\frac{\dot{A}}{A}\right)
\right\}\frac{\dot{C}}{C}+\left(\frac{\dot{B}}{B}f'_R-\dot{f'_R}+\frac{A'}
{A}\dot{f_R}\right)\right.\\\label{B1}
&\times\left.\frac{1}{A^2B^2}\left(\frac{3A'}{A}+\frac{B'}{B}+\frac{2C'}
{C}\right)\right],
\\\nonumber
D_1&=\frac{1}{\kappa}\left[\left\{\frac{1}{B^2A^2}\left(\frac{A'}{A}
\dot{f_R}-\dot{f'_R}+\frac{\dot{B}}{B}f'_R\right)\right\}_{,0}
+\frac{1}{B^2}\left\{\frac{\ddot{f_R}}{A^2}-\frac{f_R}{2}\left(
\frac{f}{f_R}\right.\right.\right.\\\nonumber
&\left.\left.\left.-R\right)-\frac{f'_R}{B^2}\left(\frac{A'}{A}
+2\frac{C'}{C}\right)-\frac{\dot{f_R}}{A^2}\left(\frac{\dot{A}}{A}
-2\frac{\dot{C}}{C}\right)\right\}_{,1}+\frac{1}{B^2}\left\{\frac{
f''_R}{B^2}+\frac{\ddot{f_R}}{A^2}\right.\right.\\\nonumber
&\left.\left.-\left(\frac{A'}{A}+\frac{B'}{B}\right)\frac{f'_R}{B^2}
-\left(\frac{\dot{A}}{A}+\frac{\dot{B}}{B}\right)\frac{\dot{f_R}}{A^2}
\right\}\frac{A'}{A}+\frac{2}{B^2}\left\{-\frac{f'_R}{B^2}
\left(\frac{B'}{B}+\frac{C'}{C}\right)\right.\right.\\\nonumber
&\left.\left.+\frac{f''_R}{B^2}-\frac{\dot{f_R}}{A^2}
\left(\frac{\dot{B}}{B}+\frac{3\dot{C}}{C}\right)\right\}\frac{C'}{C}
-\frac{1}{(AB)^2}\left(\dot{f'_R}-\frac{\dot{B}}{B}f'_R
-\frac{A'}{A}\dot{f_R}\right)\right.\\\label{B2}
&\left.\times\left(\frac{3\dot{B}}{B}+\frac{\dot{A}}{A}
+\frac{2\dot{C}}{C}\right)\right].
\end{align}
The quantities $\varphi_\mu,~\varphi_{P_r},~\varphi_{P_\bot}$ and
$\varphi_q$ are
\begin{align}\nonumber
\varphi_\mu&=\frac{A^2}{\kappa}\left[\frac{2{\epsilon}R''}{B^2}
+\frac{\lambda_n(n-1)(2{\epsilon}R)^n}{\epsilon(2BR)^2}
\left\{\frac{(n-2){R'}^2}{R}+{R''}\right\}-\left(\frac{
\dot{B}}{B}+2\frac{\dot{C}}{C}\right)\right.\\\nonumber
&\left.\times\left\{\frac{2{\epsilon}\dot{R}}{A^2}+\frac{\lambda_n(n-1)
(2{\epsilon}R)^n\dot{R}}{2{\epsilon}R^2A^2}\right\}-\frac{{\epsilon}R^2}{2}
+\frac{\lambda_n(n-1)(2{\epsilon}R)^n}{8n\epsilon}-\left\{\frac{2{\epsilon}
{R'}}{B^2}\right.\right.\\\label{B3}
&\left.\left.+\frac{\lambda_n(n-1)(2{\epsilon}R)^n{R'}}{2{\epsilon}
R^2B^2}\right\}\left(\frac{B'}{B}-\frac{2C'}{C}\right)\right],\\\nonumber
\varphi_{P_r}&=-\frac{B^2}{\kappa}\left[\left\{\frac{2{\epsilon}\dot{R}}{A^2}
+\frac{\lambda_n(n-1)(2{\epsilon}R)^n\dot{R}}{2{\epsilon}R^2A^2}\right\}
\left(\frac{\dot{A}}{A}-\frac{2\dot{C}}{C}\right)+\frac{\lambda_n(1-n)
(2{\epsilon}R)^n}{8{\epsilon}n}\right.\\\nonumber
&\left.-\frac{{\epsilon}R^2}{2}-\frac{2\epsilon\ddot{R}}{A^2}+\frac{
\lambda_n(1-n)(2{\epsilon}R)^n}{\epsilon(2AR)^2}\left\{\ddot{R}+\frac{
(n-2)\dot{R}^2}{R}\right\}+\left(\frac{A'}{A}+\frac{2C'}{C}\right)\right.\\\label{B4}
&\left.\times\left\{\frac{2{\epsilon}{R'}}{B^2}+\frac{\lambda_n(n-1)
(2{\epsilon}R)^n{R'}}{2{\epsilon}R^2B^2}\right\}\right],\\\nonumber
\varphi_{P_\bot}&=-\frac{C^2}{\kappa}\left[\left(\frac{\dot{A}}{A}-
\frac{\dot{B}}{B}+\frac{\dot{C}}{C}\right)\left\{\frac{2{\epsilon}
\dot{R}}{A^2}+\frac{\lambda_n(n-1)(2{\epsilon}R)^n\dot{R}}{2{\epsilon}
R^2A^2}\right\}+\frac{\lambda_n(1-n)}{8{\epsilon}n}\right.\\\nonumber
&\left.\times(2{\epsilon}R)^n+\left\{\frac{2{\epsilon}{R'}}{B^2}
+\frac{\lambda_n(n-1)(2{\epsilon}R)^n{R'}}{2{\epsilon}R^2B^2}\right\}
\left(\frac{C'}{C}-\frac{B'}{B}+\frac{A'}{A}\right)+\left(\frac{R''}{B^2}
\right.\right.\\\nonumber
&\left.\left.-\frac{\ddot{R}}{A^2}\right)2\epsilon+\frac{\lambda_n(n-1)
(2{\epsilon}R)^n}{4{\epsilon}R^2}\left\{\frac{R''}{B^2}-\frac{\ddot{R}}
{A^2}+\frac{(n-2)R'^2}{RB^2}-\frac{(n-2)\dot{R}^2}{RA^2}\right\}\right.\\\label{B5}
&\left.-\frac{{\epsilon}R^2}{2}\right],\\\nonumber
\varphi_q&=\frac{1}{\kappa}\left[\frac{\lambda_n(n-1)(2{\epsilon}R)^n}
{4{\epsilon}R^2}\left\{\frac{(n-2)\dot{R}R'}{R}+\dot{R'}\right\}-2{\epsilon}
\left(\frac{R'\dot{B}}{B}+\frac{\dot{R}A'}{A}\right)\right.\\\label{B6}
&\left.-\frac{\lambda_n(n-1)(2{\epsilon}R)^n}{2{\epsilon}R^2}
\left(\frac{R'\dot{B}}{B}+\frac{\dot{R}A'}{A}\right)+2{\epsilon}\dot{R'}\right].
\end{align}

\vspace{0.25cm}

{\bf Acknowledgment}

\vspace{0.25cm}

We would like to thank the Higher Education Commission, Islamabad,
Pakistan for its financial support through the {\it Indigenous Ph.D.
Fellowship for 5K Scholars, Phase-II, Batch-I}.

\end{document}